\documentclass[endfloats,showpacs,amsmath,amssymb,prl,superscriptaddress,twocolumn]{revtex4}

\usepackage{graphicx}
\usepackage{bm}
\begin{document}
\title{Associated Production of Neutral Higgs Boson with Squark Pair in the Minimal Supersymmetric Standard Model with Explicit CP Violation at the CERN LHC}
\author{Zhao Li}
\email{zhli.phy@pku.edu.cn}
\author{Chong Sheng Li}
\email{csli@pku.edu.cn}
\author{Qiang Li}
\email{qliphy@pku.edu.cn}
\affiliation{Department of Physics, Peking
University, Beijing 100871, China}

%%%%%%%%%%%%%%%%%%%%%%%%%%%%%%%%%%%%%%%%%%%%%%%%%%%%%%%%%%%%%%%%%%
\begin{abstract}
%%%%%%%%%%%%%%%%%%%%%%%%%%%%%%%%%%%%%%%%%%%%%%%%%%%%%%%%%%%%%%%%%
We investigate the associated production of neutral Higgs boson with
squark pair in the minimal supersymmetric standard model with and
without explicit CP violation, respectively. We show that the
dominant productions in both cases are always ones of the lightest
neutral Higgs boson associated with the lightest stop pair, which
can reach a few pb, in the ranges of parameters allowed by
constraints from the electric dipole moment experiments. In most of
the parameter space, the total cross sections in the case with
explicit CP violation are significantly enhanced, compared with ones
without explicit CP violation. For some special parameters, several
orders of magnitude enhancement can be obtained.
\end{abstract}


\maketitle

\noindent

The minimal supersymmetric standard model (MSSM) with CP
conservation (CPC) provides three neutral Higgs bosons: one CP odd
boson ($A^0$) and two CP even boson ($h^0$, $H^0$), whose mass
eigenstates are also their CP eigenstates. However, in the MSSM with
explicit CP violation (CPV), CP violation can introduce the mixing
of three neutral Higgs bosons, and CP eigenstates no longer exist.
The mixing can form three new mass eigenstates of the neutral
Higgs bosons: $h_1$, $h_2$ and $h_3$, and all couplings involved the
neutral Higgs bosons are changed\cite{pilaftsistocarena,Choi:2000wz}.
The previous studies have shown that explicit CP violation mentioned
above can significantly change the predictions for the Higgs productions
and decays \cite{pilaftsistocarena,Choi:2000wz,
demirtodedes,Cao:2005zk}. Especially, Q. H. Cao et al.\cite{Cao:2005zk}
recently found that there is a parameter region,
in which the Higgs production cross sections via gluon
fusion can be greatly enhanced with a large CP violation angle. However,
in their calculations they did not consider the constraints on
parameters from the experiments of electric dipole moments (EDMs),
because they believe that there are possible cancelations between
different contributions to EDMs.

In fact, the present EDM experiments
\cite{Harris:1999jx,Regan:2002ta,Romalis:2000mg,chotodemille} have
put some constraints on the parameter space in the MSSM with
explicit CP violation in some scenarios. Therefore, it is necessary
to make further investigation on the production of Higgs boson with
a large CP violation angle, using the EDMs constraints on the relevant
parameters. B- and K-physics may provide the constraints on
parameters, but they depend strongly on the flavor structure in
supersymmetry breaking\cite{Cao:2005zk}. Therefore, the constraints
potentially arising from B- and K-physics are not considered in this
letter.

In this letter, we first consider the EDM experiment constraints on
the relevant parameters in the case of CPV, and then using the
allowed parameters, we investigate the neutral Higgs production
associated with squark pair in the MSSM with and without explicit CP
violation at the CERN Large Hadron Collider (LHC)
\cite{changtoolive,Pilaftsis:1999td,Demir:2003js,
Olive:2005ru,Ibrahim:1997gj,Ibrahim:1998je}. A similar process, the
production of Higgs associated with squark pairs, has been
calculated in the minimal supergravity (mSUGRA) without CPV in
Ref.\cite{Dedes:1999ku}.

Through the one-loop effective potential of the neutral Higgs
sector, as shown in Ref.\cite{Choi:2000wz}, the complex parameters
in the mass matrices of sfermions can induce CPV in the Higgs
sector. The input parameters of the mass matrix can be
chosen as $m_A^2$, $A_t\mu\rm e^{i\xi}$ and $A_b\mu\rm e^{i\xi}$,
and there are only two CP violation angles:
$\kappa_1=\theta_{A_t}+\theta_\mu+\xi$ and
$\kappa_2=\theta_{A_b}+\theta_\mu+\xi$. Therefore, it is equivalent
that we choose $\theta_\mu=\xi=0$, and leave $\theta_{A_t}$ and
$\theta_{A_b}$ as the CP violation angles. Diagonalizing the
mass matrix of neutral Higgs, we get the mass eigenstates $h_1$,
$h_2$ and $h_3$, and the corresponding mass eigenvalues $m_{h_1}$,
$m_{h_2}$ and $m_{h_3}$ ($m_{h_1}\leq m_{h_2}\leq m_{h_3}$), which
are different from the CP eigenstates, ($h^0$, $H^0$, $A^0$).
Using the relations between ($h^0,H^0$) and ($\phi_1,\phi_2$), we can
obtain the relations between ($h^0$, $H^0$, $A^0$) and ($h_1$, $h_2$, $h_3$).
The calculations of relevant processes considered
here at the leading order are straightforward, so we do not show
their expressions.

In our numerical calculations, the Standard Model (SM) parameters
are chosen to be $\alpha_{ew}(m_W)=1/137$, $m_Z=91.1875$~GeV,
$m_W=80.45$~GeV and $m_t=174$~GeV\cite{unknown:2005pe}. We use the
CTEQ6M parton distribution functions (PDFs)\cite{Pumplin:2002vw} and
the two-loop evaluation for $\alpha_s(Q)$\cite{gorishniitodjouadi}
$(\alpha_s(m_Z)=0.118)$ and the running bottom quark
mass\cite{djouaditospira}. The other SM input parameters
are\cite{Eidelman:2004wy,unknown:2005pe}
\begin{equation*}
\begin{array}{lll}
m_e=0.51MeV, & m_\mu=105.658MeV, & m_\tau=1.777GeV,\\
m_d=4MeV, & m_s=120MeV, & m_b=4.2GeV, \\
m_u=2.5MeV, & m_c=1.2GeV. &
\end{array}
\end{equation*}
In addition, we assume that the complex parameters only exist in the
squark sector. For simplicity, we choose
$A_t=A_b=A_u=A_d=Ae^{i\theta_A}$ and $m_{\tilde Q}=m_{\tilde
U}=m_{\tilde D}=m_{\tilde E}=m_{\tilde L}=M_{SUSY}$. Then, using the
solution of $m_{\tilde t_1}$, we can choose $m_{\tilde t_1}$ as
input parameter instead of $M_{SUSY}$. The other MSSM input
parameters are fixed as following:
\begin{equation*}
\begin{split}
& M_1=100GeV,~~~M_2=200GeV,~~~M_3=300GeV, \\
& A_l=A_\tau=1000GeV,~~~\theta_A=\pi/2,~~~\theta_\mu=\xi=0.
\end{split}
\end{equation*}
The remaining MSSM input parameters are $\tan\beta$, $M_{A^0}$,
$m_{\tilde t_1}$, $A$ and $\mu$.

The strongest constraints on the parameters in the case of CPV arise
from the experiments on the EDMs of thallium\cite{Regan:2002ta},
mercury\cite{Romalis:2000mg} and neutron\cite{Harris:1999jx}.
Although it is supposed that cancelations possibly come from many
contributions \cite{ellistopokorski, Ibrahim:1997gj,Ibrahim:1998je,
Cao:2005zk}, it is still necessary to investigate further the
constraints on the parameter space in the case of explicit CP
violation based on the recent detailed theoretical analyses
\cite{changtoolive,Pilaftsis:1999td,Ibrahim:1998je,Ibrahim:1997gj,Demir:2003js,Olive:2005ru}.
The current experiments at LEPII and Tevatron have also set up the
lower bounds\cite{Eidelman:2004wy,unknown:2005pe} on the masses of
neutral Higgs, stop and sbottom, which are $89.8$~GeV, $95.7$~GeV
and $89$~GeV, respectively, and $|A_t|$ and $|A_b|$ can not be taken
too large\cite{Dedes:1999ku,kounnastocasas} in order to avoid a
color breaking vacuum expectation value (VEV).

Combining all the above constraints, we can obtain the allowed
regions in the parameter space.

From the EDM expressions, we find that parameters spaces
are constrained more stringently when $M_{A^0}$ becomes smaller. In
order to calculate numerical results in larger ranges of parameters
allowed by constraints from all experiments, we will consider the
case of $M_{A^0}=300$~GeV below.

\begin{figure}[ht!]
\includegraphics[width=0.42\textwidth]{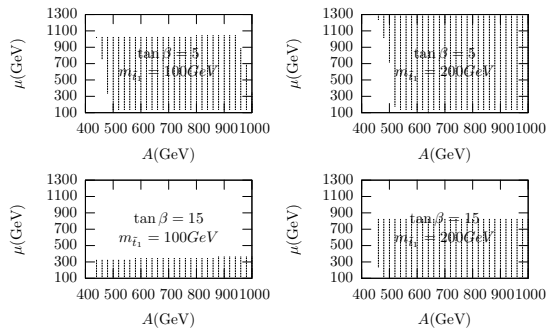}
\caption{The allowed ranges of the parameters $A$ and $\mu$ for
$M_{A^0}=300$GeV, $\tan\beta=5$ and $15$,
$m_{\tilde t_1}=100$GeV and $200$GeV, respectively.
The shaded areas correspond to the allowed ranges of parameters.}
\label{edm2}
\end{figure}

\begin{figure}[ht!]
\includegraphics[width=0.42\textwidth]{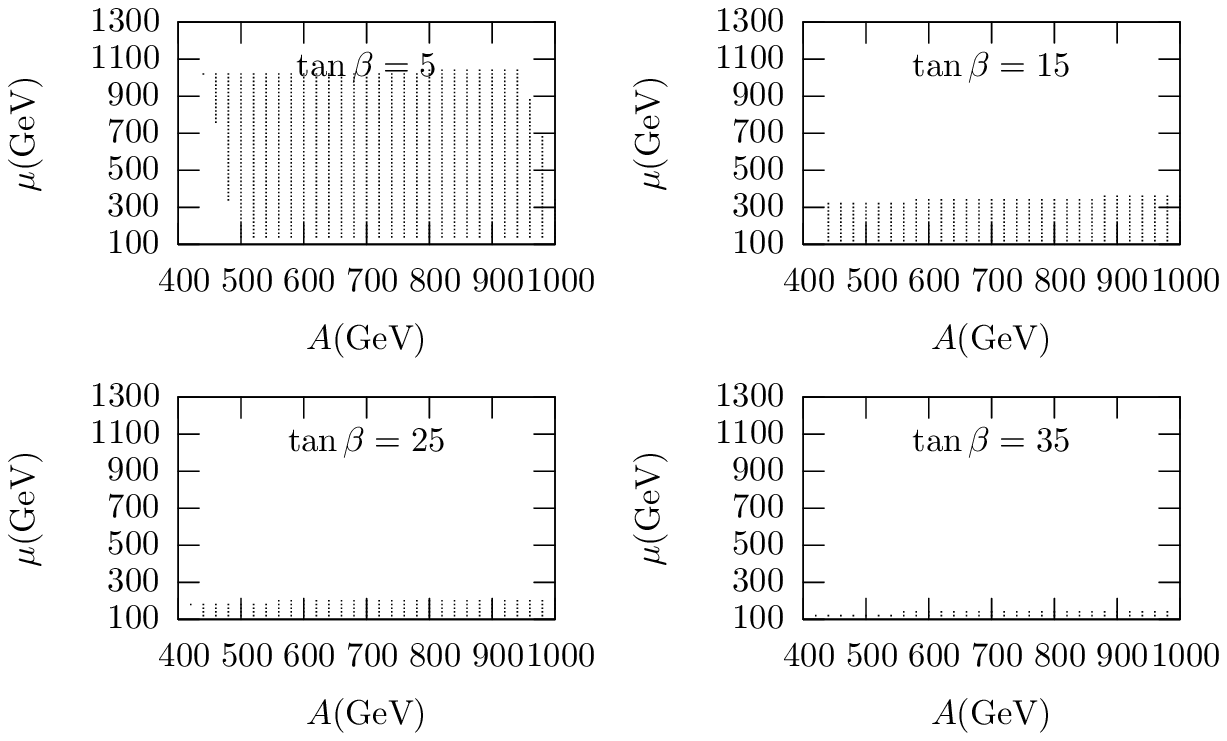}
\caption{The allowed ranges of the parameters $A$ and $\mu$ for
$M_{A^0}=300$GeV, $m_{\tilde t_1}=100$GeV,
$\tan\beta=5$, $15$, $25$ and $35$, respectively.
The shaded areas correspond to the allowed ranges of parameters.}
\label{edm3}
\end{figure}

Figs.\ref{edm2}-\ref{edm3} show the allowed ranges of $A$ and $\mu$
for different $\tan\beta$ and $m_{\tilde t_1}$. In general, the
constraints on both $A$ and $\mu$ become weaker for larger
$m_{\tilde t_1}$, and the allowed ranges of $\mu$ are sharply
constrained for large $\tan\beta$, especially for $\tan\beta=35$. We
will mainly consider the case of $\tan\beta\geq5$ in the following.

\begin{figure}[ht!]
\includegraphics[width=0.42\textwidth]{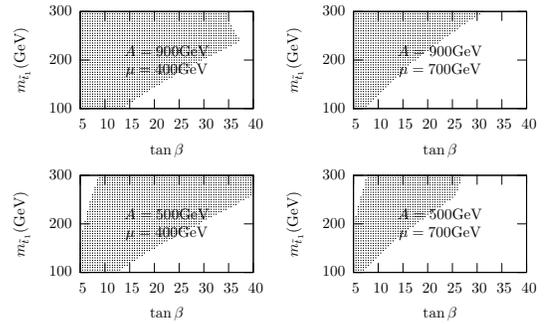}
\caption{The allowed ranges of the parameters $\tan\beta$ and $m_{\tilde t_1}$ for
$M_{A^0}=300$GeV, $A=500$GeV and $900$GeV,
$\mu=400$GeV and $700$GeV, respectively.
The shaded areas correspond to the allowed ranges of parameters.}
\label{edm4}
\end{figure}

Fig.\ref{edm4} gives the allowed ranges of $\tan\beta$ and
$m_{\tilde t_1}$ for different $A$ and $\mu$. $\tan\beta$ can be
strongly constrained for larger $\mu$ and smaller $A$. For example,
for $A=500$~GeV and $\mu=700$~GeV, $\tan\beta$ is almost limited
near $5$, and in the case of $\mu=700$~GeV, $m_{\tilde t_1}$ is
strongly constrained only for large $\tan\beta$.

Using the allowed ranges mentioned above, we investigate the
productions of neutral Higgs associated with squark pair in the MSSM
with and without explicit CP violation, respectively. In the
experiments, the mass eigenstates of Higgs are the observables,
which can be used to distinguish different Higgs bosons, and for
comparison with the case of CPC, we will mainly focus on the
productions of the lightest Higgs associated with squark pairs. In
the following, we define $\Phi=(h_1,h_2,h_3)$ for the case of CPV,
and $\Phi=(h^0,H^0,A^0)$ for the case of CPC, respectively.

\begin{figure}[ht!]
\includegraphics[width=0.42\textwidth]{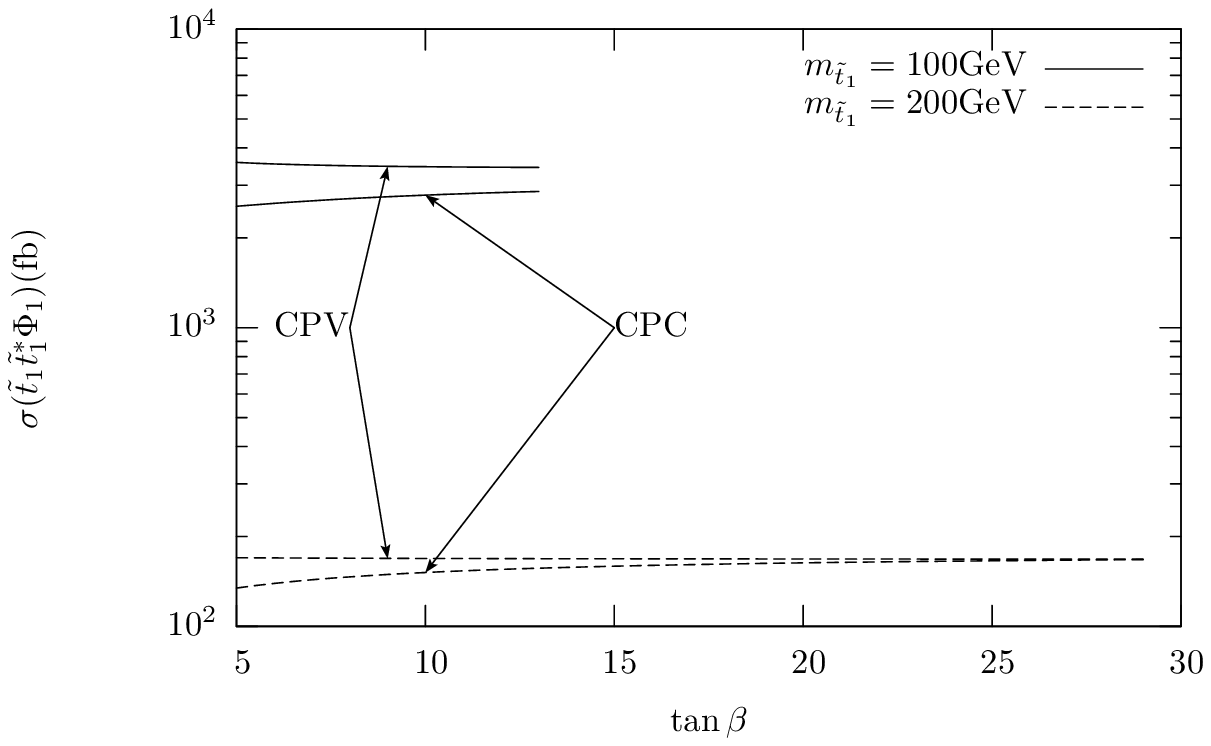}
\includegraphics[width=0.42\textwidth]{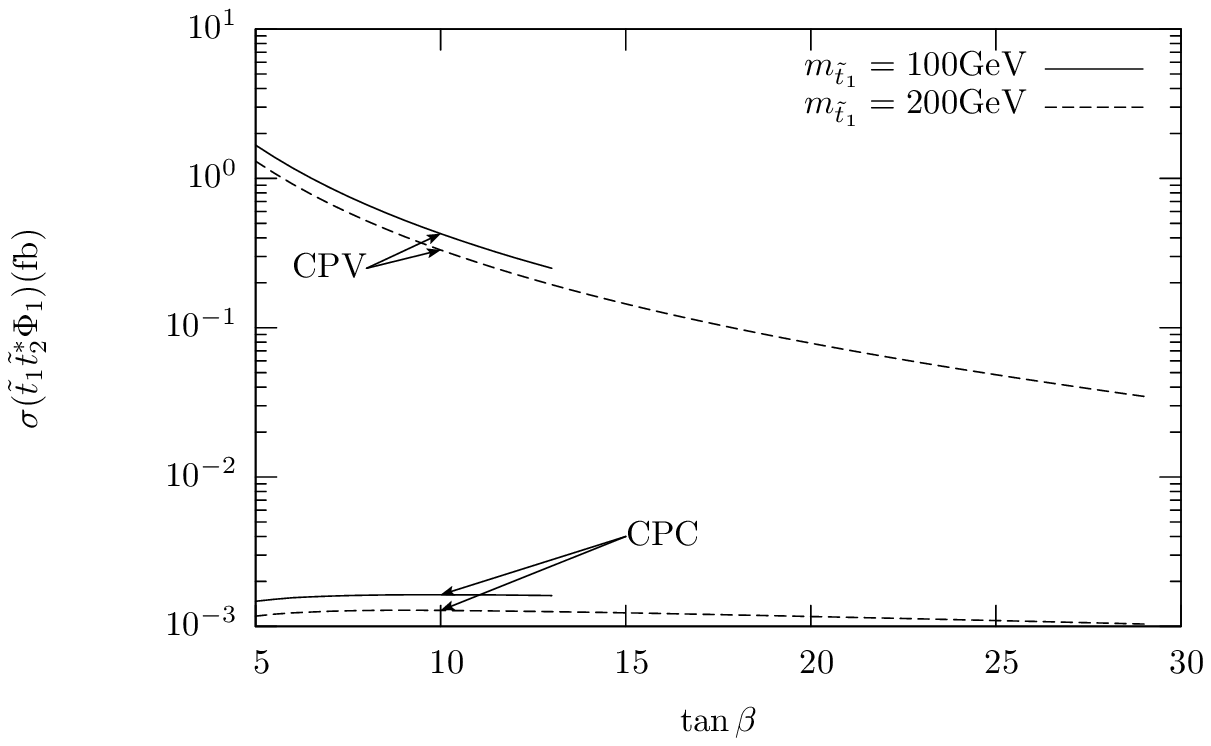}
\caption{The cross sections of $\Phi_1$ productions associated with squark pairs
as the functions of $\tan\beta$, where $\Phi_1=h^0$ in CPC and $\Phi_1=h_1$ in CPV,
assuming $A=900$~GeV and $\mu=400$~GeV.}
\label{h1rantb}
\end{figure}

Fig.\ref{h1rantb} shows the total cross sections of two processes
$pp\to\Phi_1\tilde t_1\tilde t_1^*$ ($\Phi_1=h_1,h^0$) and
$pp\to\Phi_1\tilde t_1\tilde t_2^*$ as functions of $\tan\beta$,
respectively, assuming $A=900$~GeV, $\mu=400$~GeV, $m_{\tilde
t_1}=100$~GeV or $200$~GeV. In both processes, the total cross
sections in the case of CPV are larger than ones in the case of CPC,
especially for $\tan\beta\approx 5$. The total cross sections for
$\Phi_1\tilde t_1\tilde t_1^*$ productions are the largest, which
are about
\begin{equation}
\sigma(h^0\tilde t_1\tilde t_1^*)\approx
2.55pb~~~and~~~\sigma(h_1\tilde t_1\tilde t_1^*)\approx 3.78pb,
\end{equation}
respectively, for $A=900$~GeV, $\mu=400$~GeV, $m_{\tilde
t_1}=100$~GeV and $\tan\beta=5$. Note that $\sigma(h_1\tilde
t_1\tilde t_1^*)$ are enhanced by about $50\%$, compared with
$\sigma(h^0\tilde t_1\tilde t_1^*)$. At the LHC, with an integrated
luminosity of 100 $fb^{-1}$ , we can expect that the events will be
about $2.55\times10^5$ for $\sigma(h^0\tilde t_1\tilde t_1^*)$ and
$3.78\times10^5$ for $\sigma(h_1\tilde t_1\tilde t_1^*)$,
respectively, and those signals may be observable in future
experiments. Fig.\ref{h1rantb} also shows that all the total cross
sections for $m_{\tilde t_1}=200$~GeV are less than ones for
$m_{\tilde t_1}=100$~GeV.

\begin{figure}[ht!]
\includegraphics[width=0.42\textwidth]{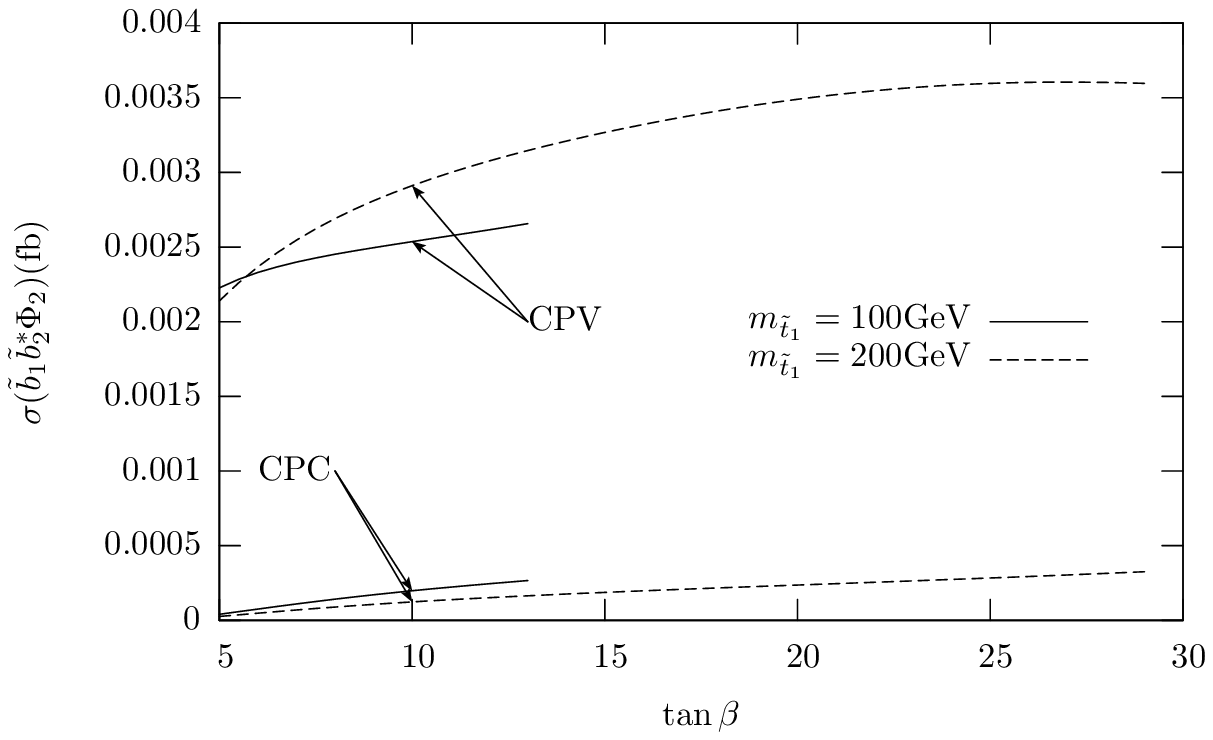}
\includegraphics[width=0.42\textwidth]{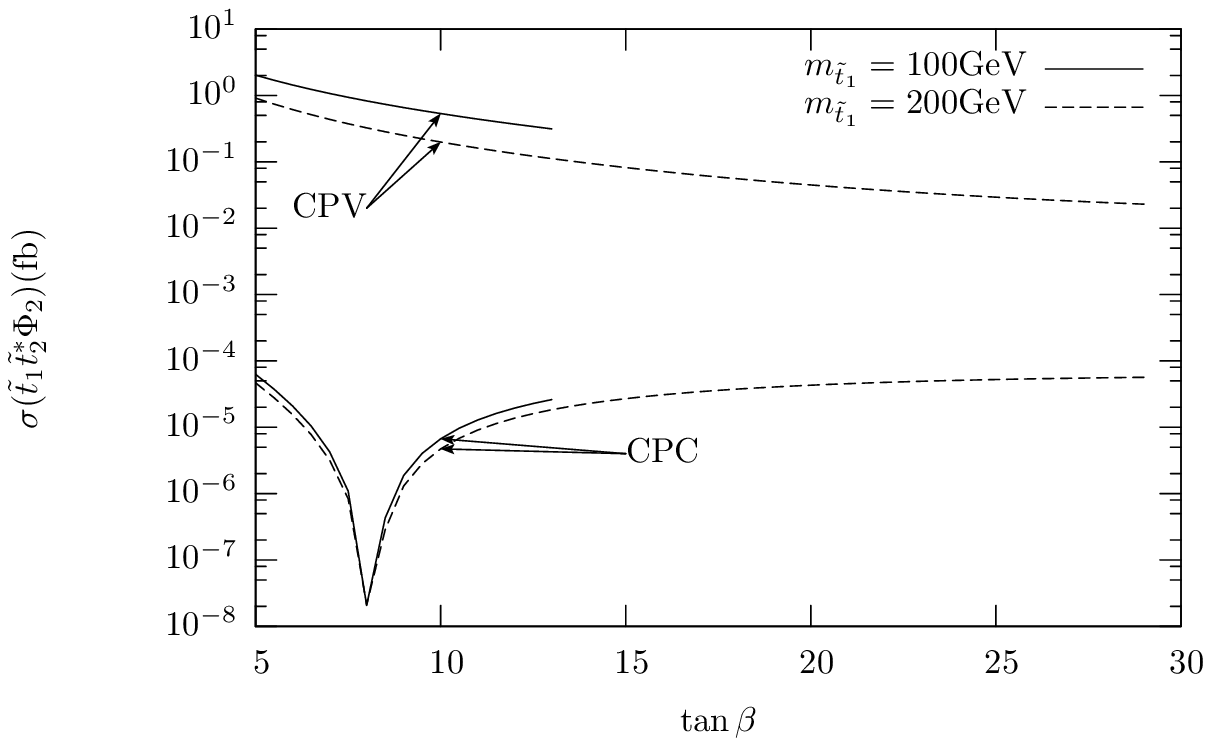}
\caption{The cross sections of $\Phi_2$ productions associated with squark pairs
as the functions of $\tan\beta$, where $\Phi_2=H^0$ in CPC and $\Phi_2=h_2$ in CPV,
assuming $A=900$~GeV and $\mu=400$~GeV.}
\label{h2rantb}
\end{figure}

In Fig.\ref{h2rantb}, we plot the total cross sections of two
processes $pp\to\Phi_2\tilde t_1\tilde t_2^*$ ($\Phi_2=h_2,H^0$) and
$pp\to\Phi_2\tilde b_1\tilde b_2^*$ as functions of $\tan\beta$,
respectively, assuming $A=900$~GeV, $\mu=400$~GeV, $m_{\tilde
t_1}=100$~GeV or $200$~GeV. Here we find that the total cross
sections for $H^0\tilde t_1\tilde t_1^*$ production are larger than
ones for $h_2\tilde t_1\tilde t_1^*$ production, and both can
generally reach tens of fb.

For the associated productions of $A^0\tilde q_i\tilde q_j^*$ and
$h_3\tilde q_i\tilde q_j^*$, the cross sections $\sigma(h_3\tilde
t_1\tilde t_1^*)$, $\sigma(h_3\tilde t_1\tilde t_2^*)$ and
$\sigma(A^0\tilde t_1\tilde t_2^*)$ are all a few fb in general, and
the other ones are all less than 1fb.

\begin{figure}[ht!]
\includegraphics[width=0.42\textwidth]{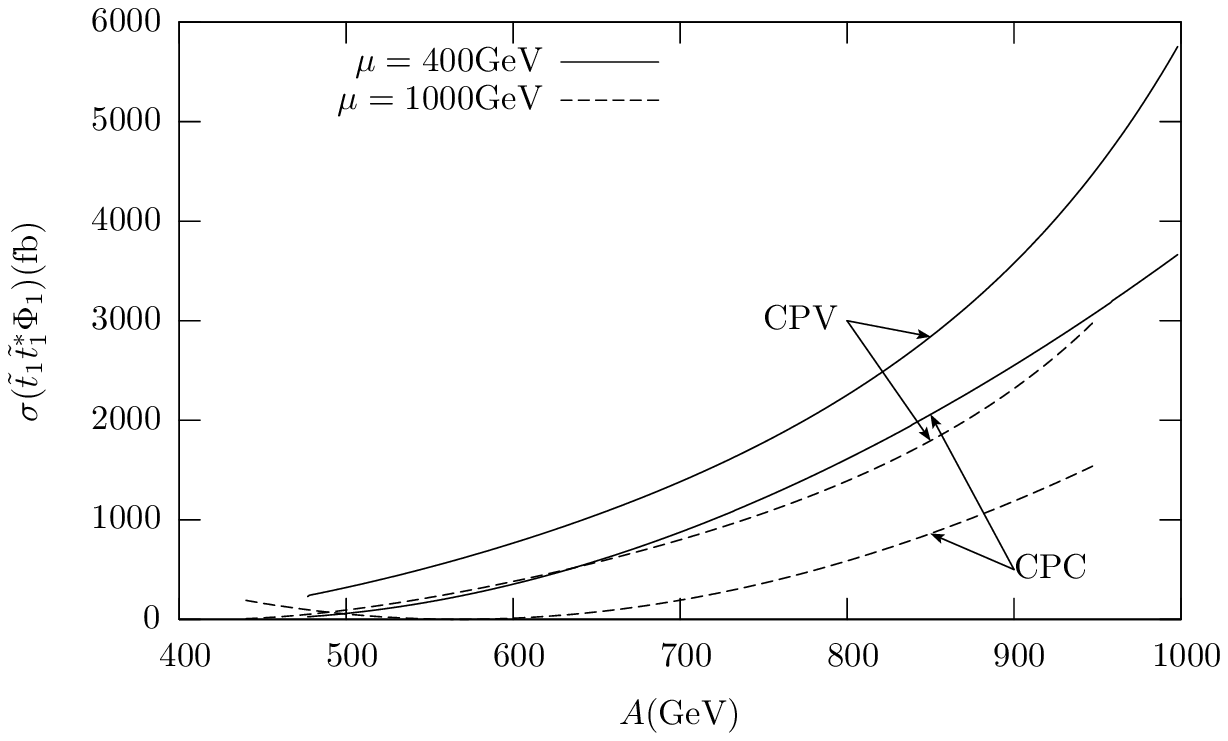}
\includegraphics[width=0.42\textwidth]{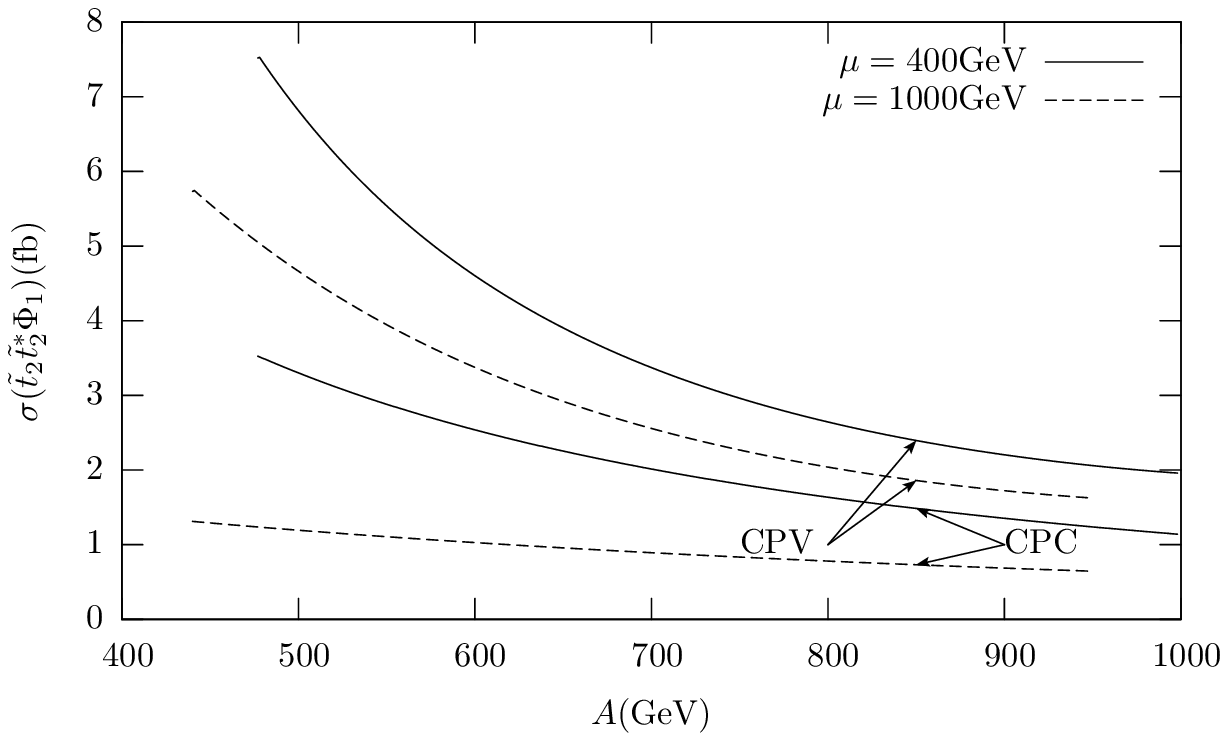}
\caption{The cross sections of $\Phi_1$ productions associated with squark pairs
as the functions of $A$, where $\Phi_1=h^0$ in CPC and $\Phi_1=h_1$ in CPV,
assuming $\tan\beta=5$ and $m_{\tilde t_1}=100$~GeV.}
\label{h1rana}
\end{figure}

Fig.\ref{h1rana} shows the curves of $\sigma(\Phi_1\tilde t_1\tilde
t_1^*)$ and $\sigma(\Phi_1\tilde t_2\tilde t_2^*)$ as functions of
$A$, respectively, assuming $\tan\beta=5$, $m_{\tilde t_1}=100$~GeV
and $\mu=400$~GeV or $1000$~GeV. The dominant processes are still
$pp\to\Phi_1\tilde t_1\tilde t_1^*$. The total cross sections
$\sigma(h_1\tilde t_1\tilde t_1^*)$ are enhanced, compared with
$\sigma(h^0\tilde t_1\tilde t_1^*)$. And when $A>700$~GeV, such
enhancement increases with increasing $A$. Especially, the
enhancement can approach to infinite for some special values of $A$
and $\mu$, because the total cross sections $\sigma(h^0\tilde
t_1\tilde t_1^*)$ approach to zero then, because the squark-Higgs
couplings tend to zero in the case of CPC. For example,
\begin{equation}
\sigma(h^0\tilde t_1\tilde t_1^*)< 10^{-3}fb,~~~\sigma(h_1\tilde
t_1\tilde t_1^*)\approx 270 fb,
\end{equation}
for $A=568$~GeV and $\mu=1000$~GeV.

\begin{figure}[ht!]
\includegraphics[width=0.42\textwidth]{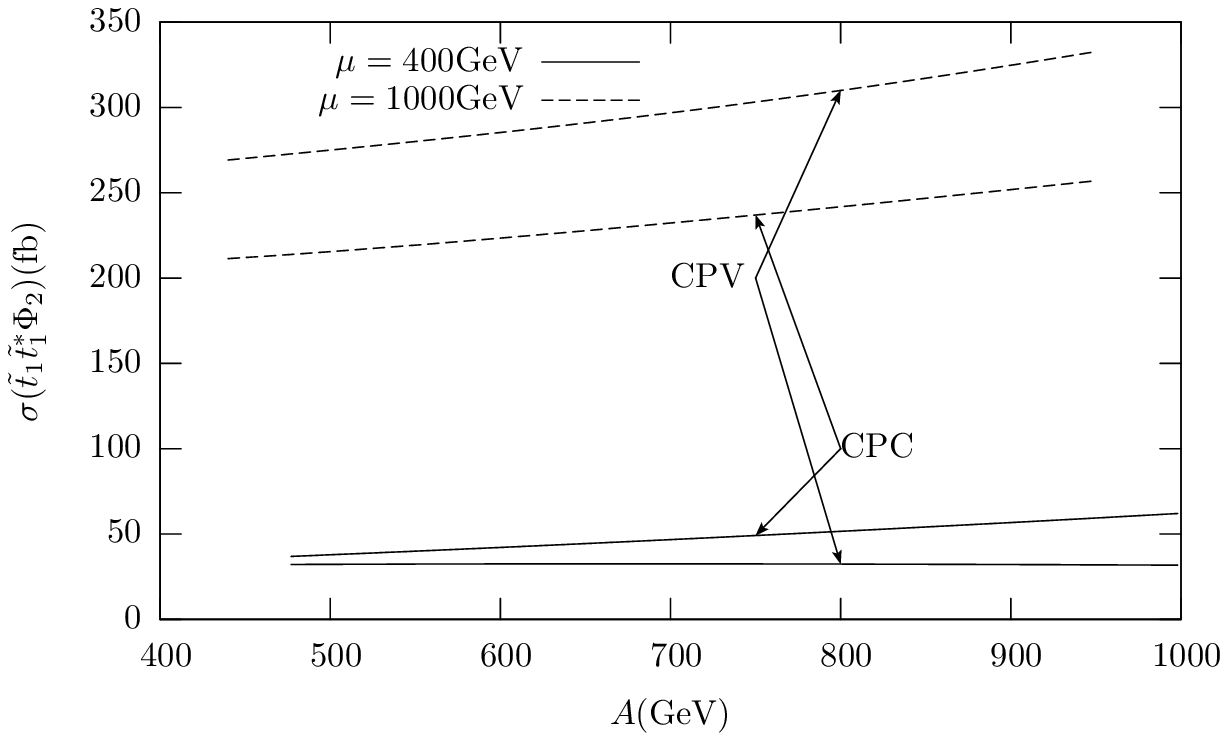}
\includegraphics[width=0.42\textwidth]{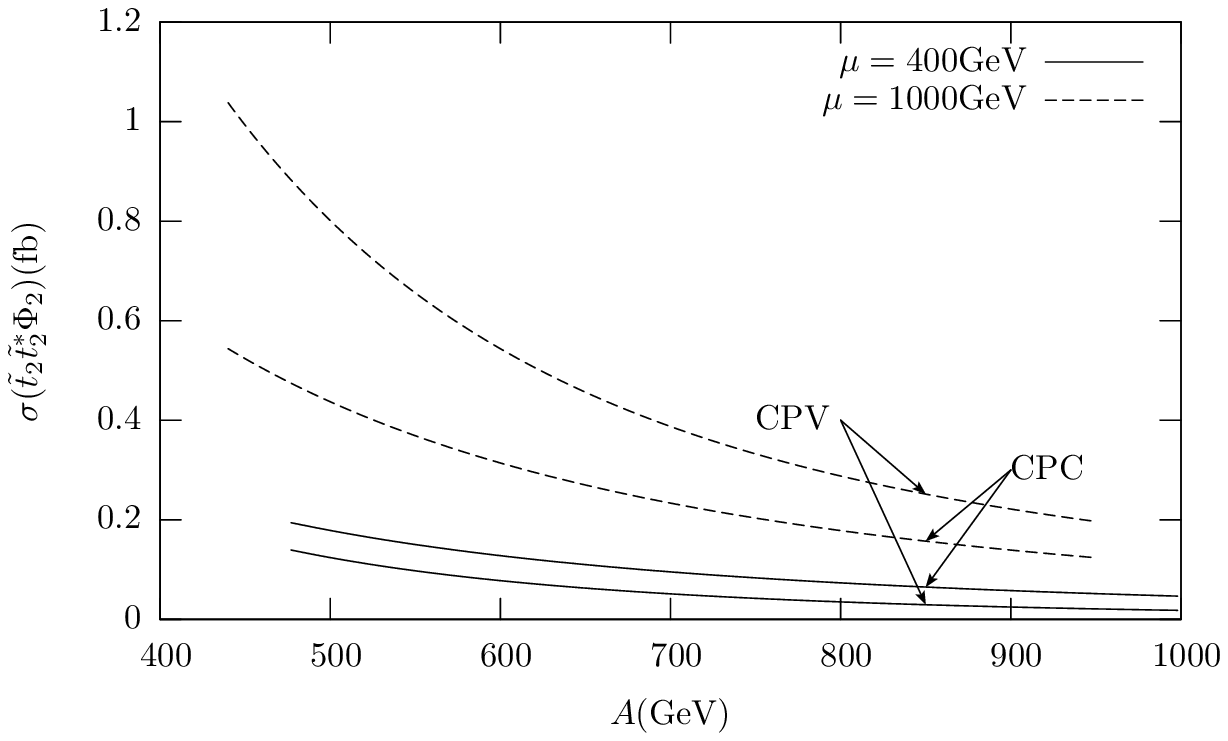}
\caption{The cross sections of $\Phi_2$ productions associated with squark pairs
as the functions of $A$, where $\Phi_2=H^0$ in CPC and $\Phi_2=h_2$ in CPV,
assuming $\tan\beta=5$ and $m_{\tilde t_1}=100$~GeV.}
\label{h2rana}
\end{figure}

Fig.\ref{h2rana} gives the total cross sections of two processes
$pp\to\Phi_2\tilde t_1\tilde t_1^*$ and $pp\to\Phi_2\tilde t_2\tilde
t_2^*$ as functions of $A$, respectively, assuming $\tan\beta=5$,
$m_{\tilde t_1}=100$~GeV and $\mu=400$~GeV or $1000$~GeV. The total
cross sections $\sigma(\Phi_2\tilde t_1\tilde t_1^*)$ can reach
hundreds of fb, corresponding to about $10^4$ events at the LHC,
while the other ones are generally less than 1fb.

For the processes $pp\to\Phi_3\tilde q_i\tilde q_j^*$
($\Phi_3=h_3,A^0)$, the total cross sections of $pp\to\Phi_3\tilde
t_1\tilde t_2^*$ are the largest, which can reach hundreds of fb,
but fastly decrease with increasing $A$. The other ones are less
than a few fb.

In conclusion, we have calculated the total cross sections of the
neutral Higgs production associated with squark pair in the MSSM
without and with explicit CP violation, respectively. The dominant
processes in both cases are always $pp\to\Phi_1\tilde t_1\tilde
t_1^*$, which can reach a few pb, while the maximum results in
mSUGRA without CP violation for the same final states are less than
500~fb \cite{Dedes:1999ku}. In most of the parameter space, the
total cross sections $\sigma(h_1\tilde t_1\tilde t_1^*)$ are always
significantly enhanced, compared with $\sigma(h^0\tilde t_1\tilde
t_1^*)$. For some special parameters, several orders of magnitude
enhancement can be obtained.

This work was supported in part by the National Natural Science
Foundation of China, under Grant Nos. 10421503 and 10575001, and
the key Grant Project of Chinese Ministry of Education, under
Grant No. 305001.
\bibliography{ggsqsqh}

\end{document}